# Large angle magnetization dynamics measured by time-resolved ferromagnetic resonance*

Th. Gerrits, M.L. Schneider, A.B. Kos and T.J. Silva

National Institute of Standards and Technology, Boulder, Colorado, 80305, USA

A time-resolved ferromagnetic resonance technique was used to investigate the magnetization dynamics of a 10 nm thin Permalloy film. The experiment consisted of a sequence of magnetic field pulses at a repetition rate equal to the magnetic system's resonance frequency. We compared data obtained by this technique with conventional pulsed inductive microwave magnetometry. The results for damping and frequency response obtained by these two different methods coincide in the limit of a small angle excitation. However, when applying large amplitude field pulses, the magnetization had a non-linear response. We speculate that one possible cause of the nonlinearity is related to self-amplification of incoherence, known as the Suhl instabilities.

# 1. INTRODUCTION

We report on results of magnetization dynamics initiated by large and small amplitude time-resolved ferromagnetic resonance (TR-FMR). We define TR-FMR as the excitation of a magnetic system by a sequence of magnetic field pulses with a repetition rate that is close or equal to the resonance frequency of the magnetic system under study. Here, we use an inductive detection technique, similar to that employed in a pulsed inductive microwave magnetometer (PIMM)[1] to acquire the TR-FMR signal. A sequence of magnetic field pulses (ranging from 1 to 16 pulses) was used to investigate the magnetodynamics in a thin Permalloy ($Ni_{80}Fe_{20}$) film. This sequence of pulses was generated by a commercial pulse/data pattern generator. In this way we could directly compare the TR-FMR (comparable to conventional FMR when the number of pulses is large) and the PIMM-method (which makes use of a single step magnetic field pulse). We show that TR-FMR and PIMM yield the same results for the extracted damping parameter and frequency response of the sample under investigation in the limit of small angle excitations.

Ferromagnetic resonance experiments can pump a magnetic system to a large angle precession cone when the damping of the system is low[2,3]. Figure 1 shows a simulation of the magnetization dynamics for a large amplitude field pulse sequence (dotted line). The inset in Fig. 1 shows the experimental configuration used throughout this study. The excitation field is always applied along the *y*-axis and the uniaxial anisotropy easy-axis of the magnetic sample lies along the *x*-axis. In this simulation the bias field is applied along the *x*-axis. In the experiment, however, the bias field may be applied along both, the *x*- and *y*-axis.

The repetition rate of the pulse sequence was equal to 786 MHz. The dashed and solid lines in Fig. 1 represent the two in-plane components of the magnetization for a thin Permalloy ($Ni_{80}Fe_{20}$) film, calculated from the Gilbert form of the Landau-Lifshitz macro-spin model[4]. For this simulation, we chose a bias field $H_x$ of 80 A/m (1.0 Oe), a uniaxial anisotropy field $H_k^{(2)}$ of 600 A/m (7.55 Oe) and a damping parameter $\alpha$ of 0.01. The pulse pattern had a peak-to-peak amplitude of 1.5 kA/m (19 Oe). This macro-spin simulation clearly shows a 90° phase shift between the x and y components of the magnetization indicating that the magnetization follows a full 360° in-plane rotation mode after the first cycle of the pulse sequence. This rotational mode is locked to the applied field for the duration of the remaining pulse sequence. The locking mechanism is initiated by a constant out-of-plane component of the magnetization $\vec{M}$. If an out-of-plane component of $\vec{M}$ is present, it will result in a demagnetizing field, which will increase the rotation frequency of $\vec{M}$ in the magnetic thin film. As a result, $\vec{M}$ can lock to the driving frequency even when it this driving frequency is slightly higher than the in-plane ferromagnetic resonance frequency at the applied bias field.

This mode-locking mechanism has already been suggested in the case of spin momentum transfer devices, where a spin polarized current rather than a magnetic field pulse sequence initiates a precessional mode.[5,6] We will show that locking $\vec{M}$ into a 360° rotation mode by TR-FMR is not possible in a quasi-infinite thin Permalloy film. We presume that $\vec{M}$ undergoes a highly non-linear process, comparable to Suhl instabilities.[7]

## 2. EXPERIMENT

We used a commercial pulse/pattern generator with a bandwidth of 3.3 GHz. The pattern generator was capable of generating a 32 bit burst data pattern with a maximum output voltage of 3.3 V. In order to achieve large amplitude field pulses the pulse pattern was externally amplified by 30 dBm. The minimum bit length was limited by the generator's bandwidth to 286 ps. The maximum bit length was 30 ns. By this we were able to apply a TR-FMR pulse pattern of up to 16 pulses with a 50% duty cycle and variable repetition frequency. The pulse pattern was launched into a coplanar waveguide (CPW) structure with 110 μm center conductor width, yielding a maximum in-plane magnetic field of 1.5 kA/m (19 Oe) peak-to-peak. The dotted line in Fig. 1 shows such a magnetic field pulse pattern with a repetition rate of 786 MHz, measured with a sampling oscilloscope. We measured the large angle response using time-resolved magnetization induced optical second harmonic generation (MSHG). This technique allows calibration and time-resolved measurements of the magnetization components and therefore allows extraction of the deflection angle and absolute value of $\vec{M}$.[8,9] $|\vec{M}|$ is a measure of the total magnetization pointing in one direction and thus is also directly related to the coherence of the magnetization motion of $\vec{M}$.[10] Determination of $|\vec{M}|$ therefore allows one to infer non-linearity (e.g. generation of spin waves) during large angle deflections. The system used was a Ti:sapphire laser, generating 100 fs short intense laser pulses to generate sufficient second harmonic light intensities. The repetition rate of the laser pulses is fixed to the laser's cavity length and is about 82 MHz. The repetition rate of the laser pulses at the magnetic sample was further reduced by a pulse picker to about 10 MHz. In order to achieve time-resolution, the laser clock and pattern generator were phase-locked. The pattern generator runs off an internal crystal-oscillator. Therefore, arbitrary triggering by an external source (the laser) was not possible. We used a phase-lock loop to match the oscillation frequency of the pattern generator's crystal with the repetition rate of the laser system.

## 3. RESULTS

In the following sections we will present the results obtained by conventional PIMM, small angle TR-FMR, and large angle TR-FMR. The sample was a 10 nm Permalloy film with a 5 nm Ta capping layer. The sample was DC magnetron sputtered directly onto a glass substrate. The glass substrate was etched with $Ar/O_2$ and Ar-ion prior to the deposition of the Permalloy film to improve adhesion. We deposited Permalloy at an Ar pressure of 0.53 Pa (4.0 mTorr). The Ta was sputtered at an Ar pressure of 0.67 Pa (5.0 mTorr).

### A. Small angle response

We first measured the response of the magnetic film by use of a PIMM. An external in-plane field is used to bias the sample magnetization perpendicular to the magnetic excitation pulse and parallel to the sample's easy axis. A CPW structure with a center conductor width of 220 μm generates a quasi-step field pulse to excite the sample. In response to the field pulse, the time-varying magnetization component along the field pulse direction induces a voltage in the waveguide that is measured with a sampling oscilloscope. We extracted the resonance frequencies and damping parameter values by fitting the Fourier transform of the time-domain data to a simple damped harmonic oscillator resonance function. Further details on the PIMM and data analysis technique can be found in earlier work by *Silva et al.*[1] and *Alexander et al.*[11]. The step-excitation amplitude was equal to 60 A/m (0.8 Oe). From these measurements we could determine the resonance frequency, dynamic anisotropy field, and damping parameter as a function of bias field. The dynamic anisotropy field of the sample was found to be 620 ± 20 A/m (7.7 ± 0.2 Oe), which is 160 A/m (2 Oe) higher than the sample's uniaxial anisotropy. Therefore we conclude that a rotatable anisotropy is present upon pulse excitation. This is in agreement with earlier studies by *Lopusnik et al.*[12] and *Schneider et al.*[13]. A normalized relaxation rate $2/\tau\omega_m$, where $\omega_m = \gamma\mu_0 M_s$, can be determined from the measured relaxation time $\tau$ of the free induction decay. The normalized relaxation rate is plotted in Fig. 2 as a function of frequency (down triangles). At low frequencies, the relaxation rate increases, as it has already been shown in earlier studies.[10,14-16]

The frequency response of the PIMM analysis was used to determine the system's resonance frequency for the TR-FMR measurements. We measured the sample response by TR-FMR for bias fields of 240 A/m (3 Oe), 480 A/m (6 Oe), 720 A/m (9 Oe) and 1.2 kA/m (15 Oe). Figure 3 shows the response of the thin film when excited by a sequence of 16 field pulses with a repetition rate of 1.16 GHz for a bias field of 720 A/m. The data were fit to a driven damped simple harmonic oscillator model at resonance. The initial exponential increase of the precession amplitude is determined by the damping of the system. The frequency is determined by the applied bias field plus the dynamic anisotropy field, because the excitation amplitude is small and the sample is biased parallel to its easy axis. The fit is in good agreement with the data and yields a damping that can be expressed in terms of a Gilbert damping parameter $\alpha = 0.0094 \pm 0.0005$ and a dynamic anisotropy field $H_k$ of 590 ± 10 A/m (7.4 ± 0.1 Oe). The measured anisotropy is in good agreement with the values obtained by conventional PIMM. Additional values of $H_k$ and $\alpha$ for all four bias fields are given in table I.

We also investigated the whether the number of generated field pulses will have any effect on the relaxation rate of the system when the pulse sequence is turned off. We used three different pulse amplitudes and pulse sequences ranging from 1 pulse up to a sequence of 16 pulses. The three pulse amplitudes $H_p$ were 24 A/m (0.3 Oe), 42 A/m (0.5 Oe) and 60 A/m (0.8 Oe), calculated using the Karlqvist equation.[17] The extracted relaxation rate did not show any significant change when the number of pulses was changed. In addition, the relaxation rates coincide for the three different pulse amplitudes, shown by the open symbols in Fig. 2. Figure 2 also shows that the relaxation rates obtained for both, the conventional PIMM and the TR-FMR technique are similar. Because the extracted relaxation rates are of the same value in the small angle limit, we may speculate that the relaxation rate will coincide for both an infinite number of pulses (conventional FMR) and PIMM measurements. Indeed, work in progress indicates that conventional FMR and PIMM measurements of damping in thin Permalloy films are in agreement.[18]

### 3. Large angle response

In the following section we will discuss the results of magnetization dynamics generated by large (1.2 kA/m) amplitude magnetic field pulse patterns. Figure 4 shows the magnetic response that results when a 1.2 kA/m (15 Oe) peak-to-peak pulse pattern with a repetition frequency of 786 MHz is applied. The pulse pattern was generated by the pattern generator and was externally amplified by 30 dBm to achieve large fields that in turn induce large precessional amplitudes. The bias field was equal to 80 A/m (1 Oe) and the data was taken using the PIMM. We cannot infer any absolute deflection angle of the magnetization from the PIMM data, as we are sensitive only to one component of the magnetization. In addition, calibration of the induced voltage to determine the actual magnitude of the magnetization component is not possible. However, analysis of the Fourier transform of the measured signal showed a peak at twice the resonance frequency. One would only expect a second peak in the Fourier transform spectrum during large angle precession since the PIMM is sensitive to motion in the *y*-direction only. Thus, we conclude that a large precession angle is present. Whether a full 360° rotation occurs can only be determined by a vector-resolved technique. In order to determine the real magnetization vector and its in-plane components we applied time- and vector-resolved magnetization induced optical second harmonic generation (MSHG).[8]

Figure 5 shows an MSHG measurement on the 10 nm Permalloy sample. The bias fields in the *x*- and *y*-direction were 220 A/m (2.75 Oe) and 0 A/m, respectively. The squares represent the deflection angle of the magnetization from the *x*-axis of the system. The open circles represent the absolute value of the magnetization $|\vec{M}|$. The dashed line in Fig. 5 shows the simulation of the magnetization response due to the pulse pattern that is shown in Fig. 1. The peak-to-peak amplitude of the pulse pattern was 1.5 kA/m (19 Oe). The simulation was done using the Gilbert form of the Landau-Lifshitz macro-spin (LLG) model. For the simulation a uniaxial anisotropy field $H_k^{(2)}$ of 440 A/m (5.5 Oe) and a rotational anisotropy field $H_k^{(0)}$ of 160 A/m (2 Oe) were used. The initial angle of the magnetization determines the angle of the anisotropy easy axis ($\theta_0$) with respect to the x-axis of the experiment and was $\theta_0 = 18°$ in our case. For the simulation, we used our best estimate of the damping parameter, $\alpha = 0.013$, based on the results for the relaxation rates, obtained from the small angle PIMM measurements, shown in Fig. 2. Using these parameters the simulation shows poor agreement with the actual data. After the first cycle of the pulse, the data shows that the magnetization seems to stay in a smaller precession cone than predicted. Furthermore, $|\vec{M}|$ slowly drops after the first cycle of the pulse to a value of about $0.85 \cdot M_s$. We found that in order to get reasonable agreement between the simulations and the MSHG measurements we had to increase the damping parameter in the simulations to a value of $\alpha = 0.050 \pm 0.005$, shown by the solid line in Fig. 5. The requirement of this very large damping parameter to fit the data suggests a non-linear response of the system when a large amplitude pulse pattern is applied and is most likely due to non-linear processes, such as incoherence of the magnetization motion.

For all bias field combinations that we tested no 360° rotation (mode) could be found. However, the damping parameter needed to fit the measured dynamics was $0.050 \pm 0.005$, independent of the applied bias field. Figure 6 shows the dynamics for the bias fields $H_x = 120$ A/m (1.5 Oe) and $H_y = 170$ A/m (2.2 Oe). The data in Fig. 7 clearly show a strong decrease in $|\vec{M}|$ to a value of about $0.75 \cdot M_s$. For this particular bias field combination one would expect a 360° in-plane rotation after the first few cycles of the pulse sequence based on our simulations using the LLG model with a damping parameter of 0.013. This particular simulation is shown in Fig. 7. The large pulse amplitude causes the magnetization to lock into a 360° rotation state in the simulation after just a few pulse cycles resulting in cumulative rotation angles larger than 360°. However, the measured data do not show such a process. Moreover, $|\vec{M}|$ drops significantly and starts to oscillate at the excitation frequency. A similar behavior occurred for other bias field combinations as can be seen in Fig. 8. The graph shows $|\vec{M}|$ as a function of time for different bias field combinations. The solid lines in Fig. 8 show the adjacent average values for $|\vec{M}|$. We determined the averaged value of $|\vec{M}|$ for a time that equals 1 ns and 7 ns (dotted lines), respectively. From that we determined the drop in $|\vec{M}|$, $\Delta m$. The values for $\Delta m$ are: 15 %, 9 %, 11 % and 4 % for the datasets shown from top to bottom. The hatched areas represent the moment in time at which the magnetization is in a locked in-plane rotation mode in the simulations. The locking moment was calculated by the LLG equation with the individual bias field combinations and a damping parameter of 0.013. The moment in time at which $\vec{M}$ is locked in the simulations coincides with the moment in time at which $|\vec{M}|$ starts dropping in the measurements. The top three datasets all exhibit a locking mode during the simulations. Note that these datasets show a large $\Delta m$ (< 15 %) during the measurements. The bottom dataset did not show any locking mode during simulation. During measurements, this dataset showed the smallest $\Delta m$ compared top all of the other datasets in Fig. 8.

We speculate that a non-linear process, similar to the Suhl instability, exists under the extreme condition of large amplitude resonant pumping. In this case, as soon as the magnetization tips out-of-plane, the demagnetizing fields associated with the out-of-plane component become the dominant contribution to the internal fields. If the magnetization is not entirely uniform out-of-plane, the discrepancy between the magnetostatic fields for two neighboring regions becomes the source for incoherent motion. This will lead to dephasing of the system, which in turn will lead to a much larger discrepancy in the out-of-plane components after the next cycle of the pulse. Therefore, the magnetization motion will become more incoherent after each pulse application.

## 4. DISCUSSION

We have shown that time-resolved ferromagnetic resonance can be used to study the dynamics of magnetic thin films. By comparing the conventional PIMM with the new TR-FMR technique, we found similar results for the extracted damping and anisotropy of a 10 nm thin Permalloy film. From that we conclude that FMR and conventional PIMM yield the same results in the small angle limit. The dynamic excitation by TR-FMR showed linearity in the small angle regime. However, an increase in the relaxation rate at low frequencies could also be found as has previously been reported for dynamics measurements. This increase at low frequencies is significant for the application to magnetic memory devices, as these are switched in the low-frequency precessional regime

At large amplitude excitations the system clearly showed a non-linear response. This non-linearity manifested itself in two ways: the reduction of $|\vec{M}|$ and an apparent increase of the relaxation rate. Even though simulations based on the LLG model indicate that one would expect $\vec{M}$ to lock in a 360° in-plane rotation mode, we were unable to generate such a mode, possibly due to non-linearities in the response of our thin film. However, a 360° in-plane rotation mode of $\vec{M}$ in thin Permalloy layers is observed in SMT devices.[5] The excitation in SMT devices happens locally (~ 50 nm). We speculate that the strong exchange interactions tend to keep $\vec{M}$ uniform over the excitation region. Hence non-uniformities in the out-of-plane components are not expected. Thus, incoherence in the magnetization will not occur as a result of self-amplified non-linearity. Therefore it seems reasonable that coherent large angle excitations may exist in SMT devices. In our case the excitation happens over a large area in the magnetic film. This allows small disturbances, with dimensions larger than the exchange length, to initiate a non-uniform internal field distribution, which in turn will be amplified by the next excitation pulse. We therefore suggest that locking $\vec{M}$ into a 360° in-plane rotation mode would be possible by TR-FMR when the magnetic film is patterned and the elements are sufficiently uniform.

The results of our study also have consequences for the precessional reversal of magnetic films.[9,19] If the magnetization distribution is not uniform over the entire sample, precessional switching can potentially still be achieved, as only a single magnetic field pulse is necessary to induce the precessional switching. In this case the Suhl instabilities can be excluded, because the non-uniformities in the out-of-plane fields are not amplified by additional field pulses. However, a small amount of incoherence will still exist upon single pulse excitation. How far this incoherence influences the final state of the reversed magnetization will depend on the initial uniformity of $\vec{M}$.

## 5. ACKNOWLEDGMENTS

This work was partially supported by a fellowship within the postdoctoral program of the German Academic Exchange Service (DAAD).

Table I

| bias field (kA/m) | $H_k$ (kA/m) | $\alpha$ |
|---|---|---|
| 0.24 | 0.62±0.01 | 0.0115±0.0005 |
| 0.48 | 0.58±0.01 | 0.0106±0.0005 |
| 0.72 | 0.59±0.01 | 0.0094±0.0005 |
| 1.20 | 0.57±0.01 | 0.0109±0.0005 |

Table I. Extracted anisotropy field and damping parameter values for different bias fields.

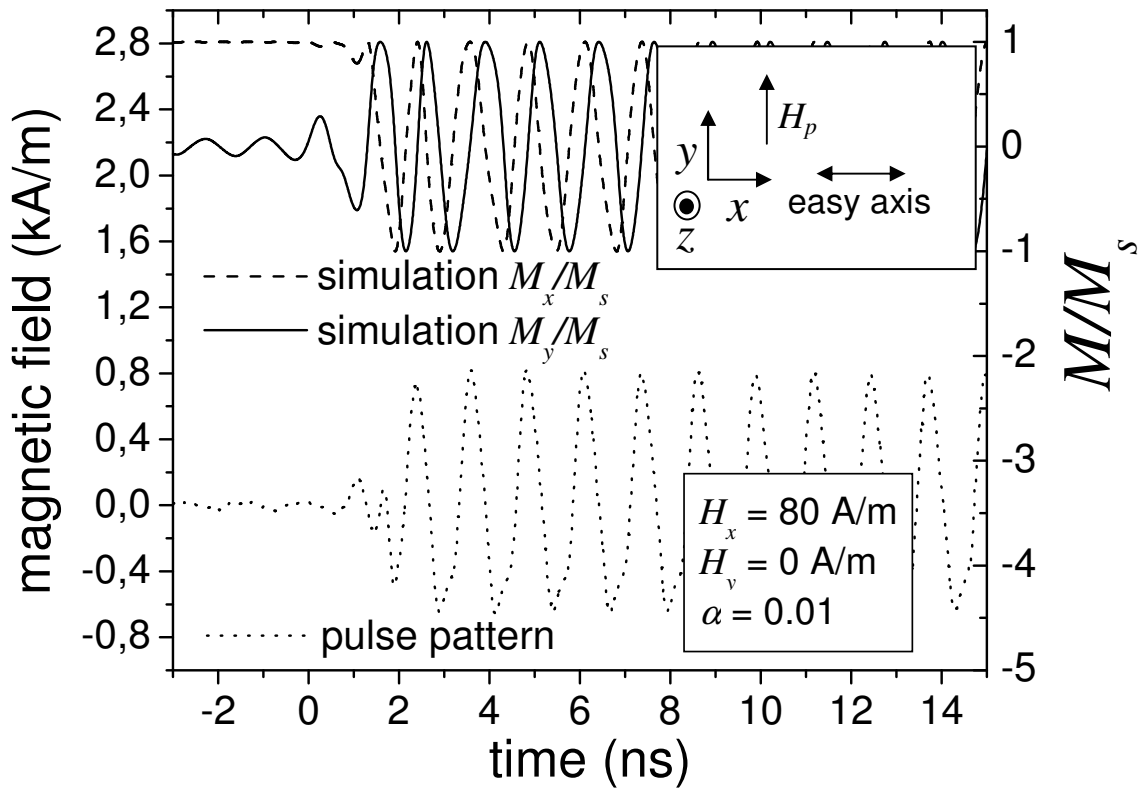

FIG 1. Landau-Lifshitz simulation of thin Permalloy film response due to a pulse field pattern (dotted line). The solid line shows the component of the magnetization along the $y$-axis and the dashed line shows the component along the $x$-axis. The parameters for the simulation are given in the lower inset. The experimental configuration is given in the upper inset. The easy axis of the sample is always parallel to the $x$-axis and the pulse field is always applied along the $y$-axis. The bias field may be applied along both, the $x$- and the $y$-axis.

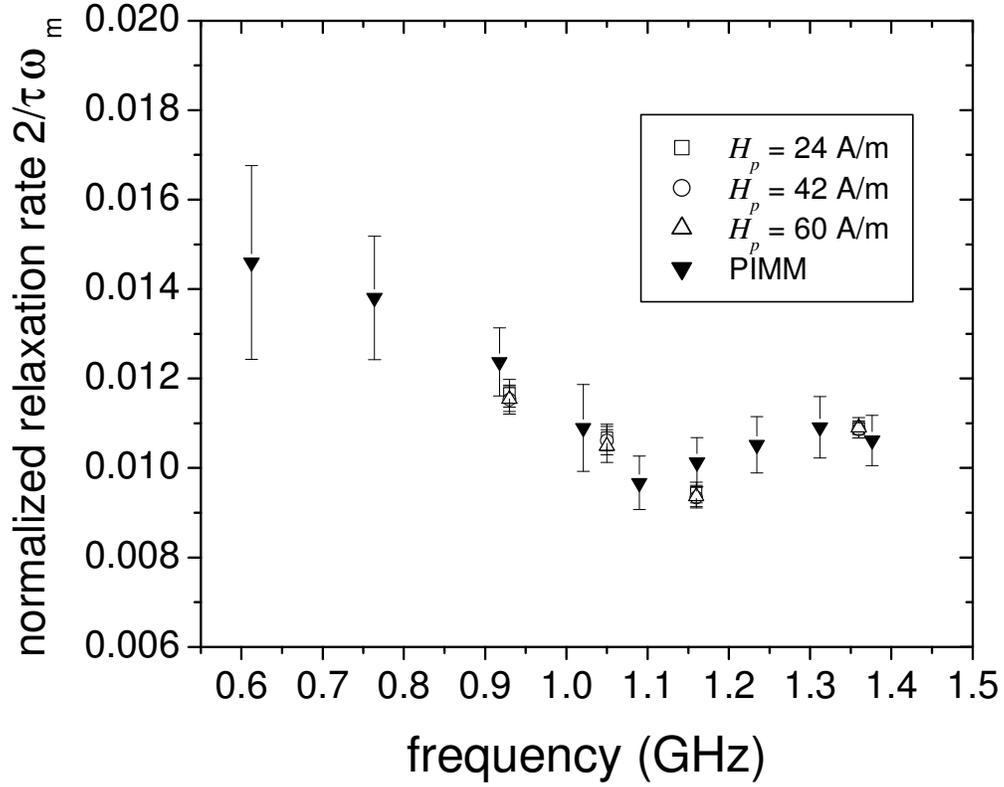

FIG 2. Extracted normalized relaxation rate as a function of frequency from PIMM-measurements (down triangles) and for TR-FMR measurements as a function of pulse amplitude $H_p$ (open symbols).

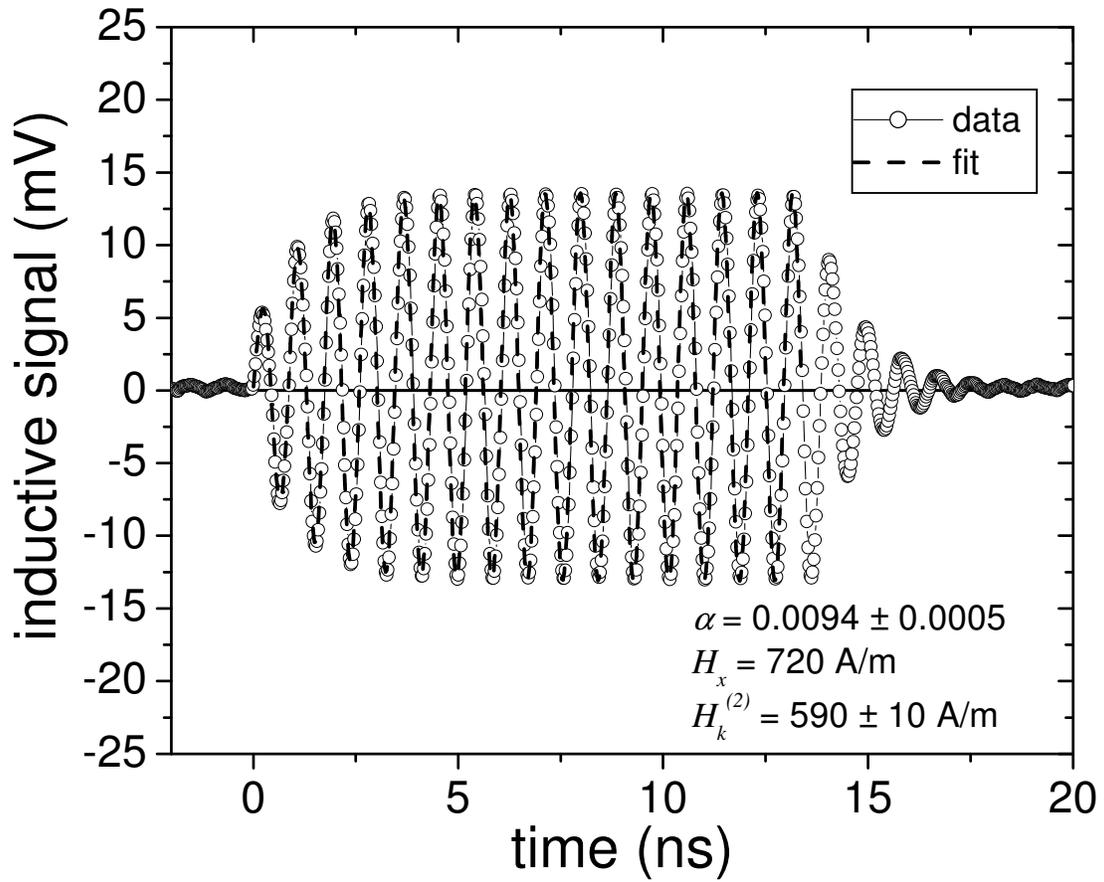

FIG 3. Time-resolved ferromagnetic resonance for a 10 nm thin Permalloy film, measured with the inductive technique. The dashed line represents the simple damped driven harmonic oscillator fit.

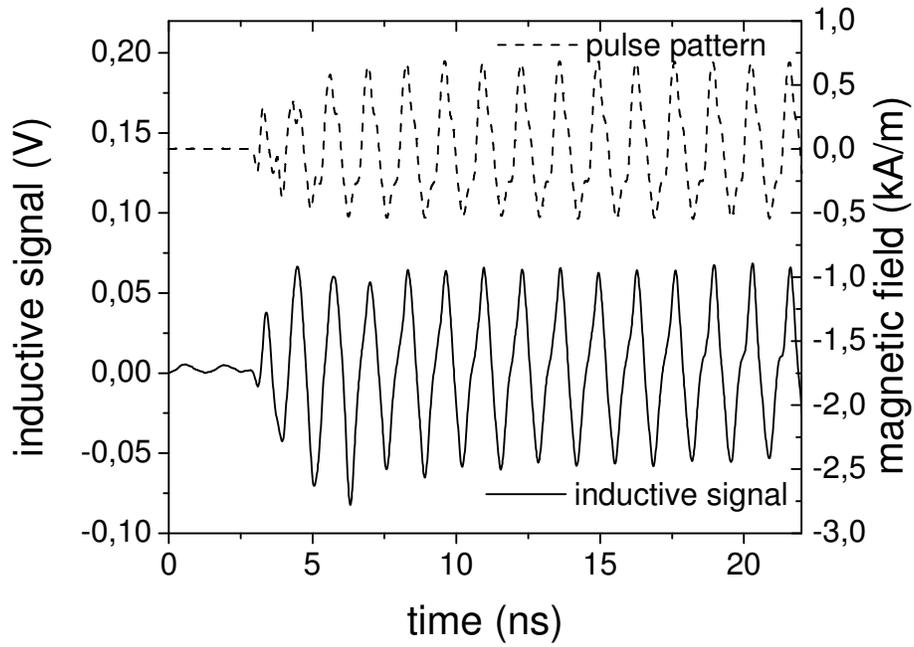

FIG 4. Inductively measured large angle excitation by large amplitude TR-FMR. The dashed line represents the pulse pattern, measured with a sampling oscilloscope. The solid line represents the inductive signal generated by the moving magnetization.

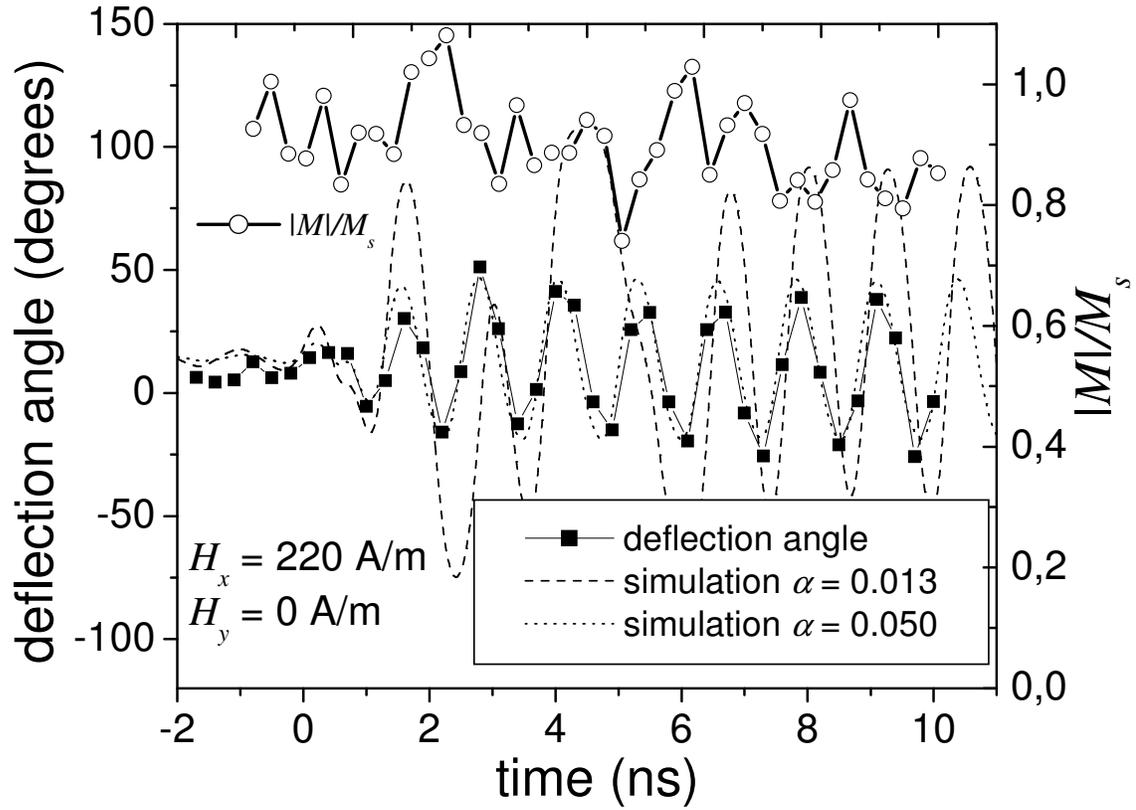

FIG 5. Vector-resolved large angle excitation measured by MSHG. The open circles represent the absolute value of the magnetization. The squares show the magnetization deflection angle from the *x*-direction. The dashed line is the LLG simulation for a damping parameter of 0.013. The dotted line shows the LLG simulation for $\alpha = 0.050$.

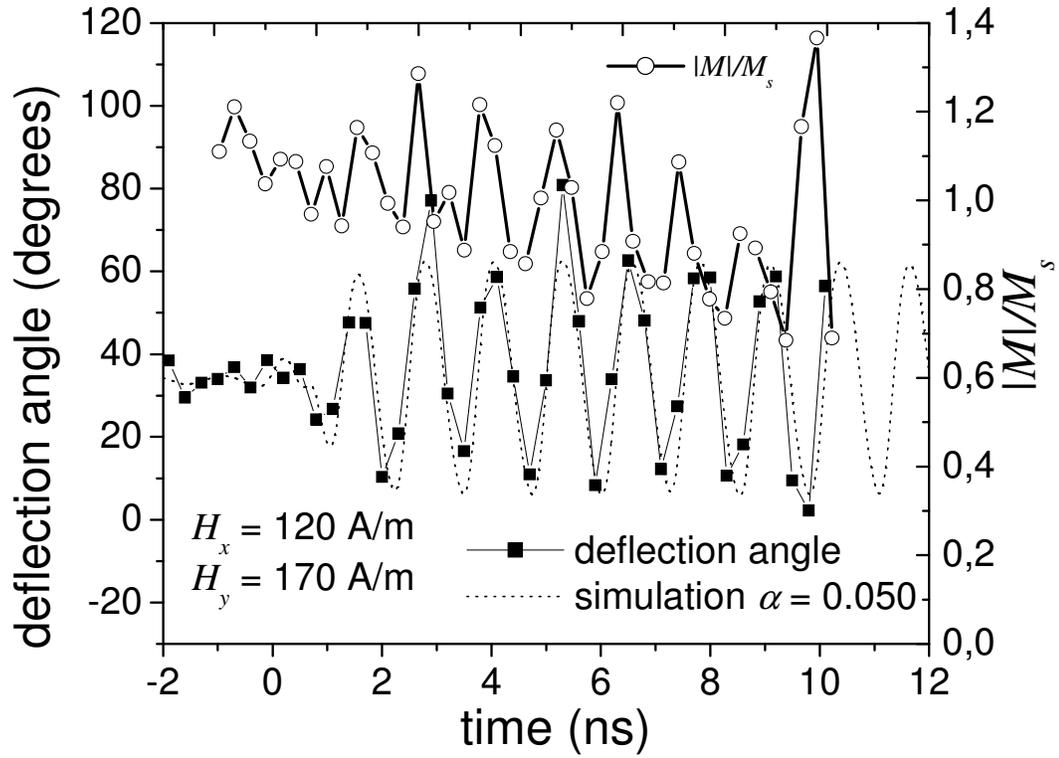

FIG 6. Vector-resolved large angle excitation measured by MSHG. The open circles represent the absolute value of the magnetization. The squares show the magnetization deflection angle from the *x*-direction. The dotted line shows the LLG simulation for $\alpha =$ 0.05.

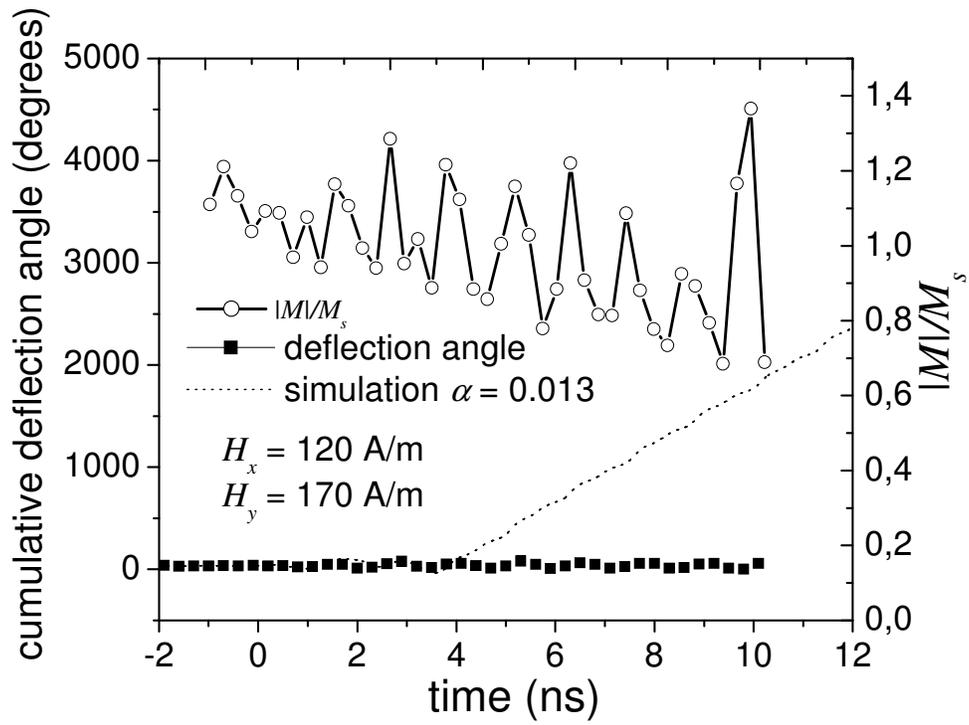

FIG 7. Vector-resolved large angle excitation measured by MSHG. The open circles represent the absolute value of the magnetization. The squares show the magnetization deflection angle from the *x*-direction. The dotted line is the LLG simulation of the deflection angle for a damping parameter of 0.013. Note that the deflection angle is cumulative and therefore exceeds values much larger than 360º.

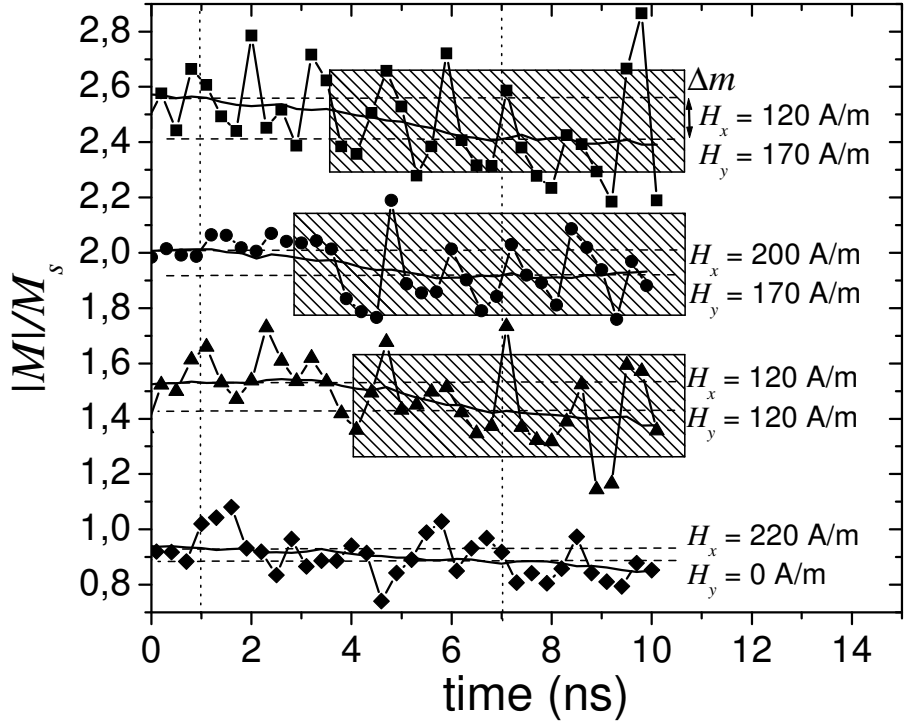

FIG 8. Closed squares, circles, triangles and diamonds show the absolute value of the magnetization for various bias field combinations. The solid lines represent the adjacent average of the data, indicating a drop in the absolute value of $\vec{M}$. The determination of $\Delta m$ was done by determination the adjacent average value of $|\vec{M}|$ at 1 ns and at 7 ns (dashed lines). The hatched areas represent the 360° locking state of the magnetization, calculated by the LLG model with a damping parameter of 0.013. Note that the upper three graphs are shifted upwards by $0.5 \cdot M/M_s$ to plot all measurements in a single graph.